\documentclass{egpubl}
\usepackage{pg2023}

\SpecialIssuePaper         % uncomment for final version of Computer Graphics Forum, special issue
\CGFStandardLicense
\usepackage[T1]{fontenc}
\usepackage{dfadobe}  

\usepackage{cite}  % comment out for biblatex with backend=biber
\BibtexOrBiblatex
\electronicVersion
\PrintedOrElectronic
\ifpdf \usepackage[pdftex]{graphicx} \pdfcompresslevel=9
\else \usepackage[dvips]{graphicx} \fi

\usepackage{egweblnk} 
\usepackage{subcaption}
\usepackage{cite}
\usepackage{float}

\title[Generating Parametric BRDFs from Natural Language Descriptions]{Generating Parametric BRDFs from Natural Language Descriptions}

\author[S. Memery \& O. Cedron \& K. Subr]
{\parbox{\textwidth}{\centering \textbf{Sean Memery\hspace{20pt}Osmar Cedron\hspace{20pt}Kartic Subr}
}
\\
{\parbox{\textwidth}{\centering 
    The University of Edinburgh \\
    Edinburgh, Scotland, UK
      }
}
}
\begin{document}

% uncomment for using teaser
% \teaser{
%  \includegraphics[width=\linewidth]{eg_new}
%  \centering
%   \caption{New EG Logo}
% \label{fig:teaser}
%}

\maketitle
%-------------------------------------------------------------------------

\begin{abstract}
Artistic authoring of 3D environments is a laborious enterprise that also requires skilled content creators. There have been impressive improvements in using machine learning to address different aspects of generating 3D content, such as generating meshes, arranging geometry, synthesizing  textures, etc. In this paper we develop a model to generate Bidirectional Reflectance Distribution Functions (BRDFs) from descriptive textual prompts. BRDFs are four dimensional probability distributions that characterize the interaction of light with surface materials. They are either represented parametrically, or by tabulating the probability density associated with every pair of incident and outgoing angles. The former lends itself to  artistic editing while the latter is used when measuring the appearance of real materials. Numerous works have focused on hypothesizing BRDF models from images of materials. We learn a mapping from textual descriptions of materials to parametric BRDFs. Our model is first trained using a semi-supervised approach before being tuned via an unsupervised scheme. Although our model is general, in this paper we specifically generate parameters for MDL materials, conditioned on natural language descriptions, within NVIDIA's Omniverse platform. This enables use cases such as real-time text prompts to change materials of objects in 3D environments such as ``dull plastic'' or ``shiny iron''. 
Since the output of our model is a parametric BRDF, rather than an image of the material, it may be used to render materials using any shape under arbitrarily specified viewing and lighting conditions.

%The tool at \url{http://dl.acm.org/ccs.cfm} can be used to generate
% CCS codes.


\end{abstract}  

\captionsetup{labelfont=bf,textfont=it}

%-------------------------------------------------------------------------
\begin{figure*}[!htbp]
\centering 
{
\includegraphics[width=1.0\linewidth]{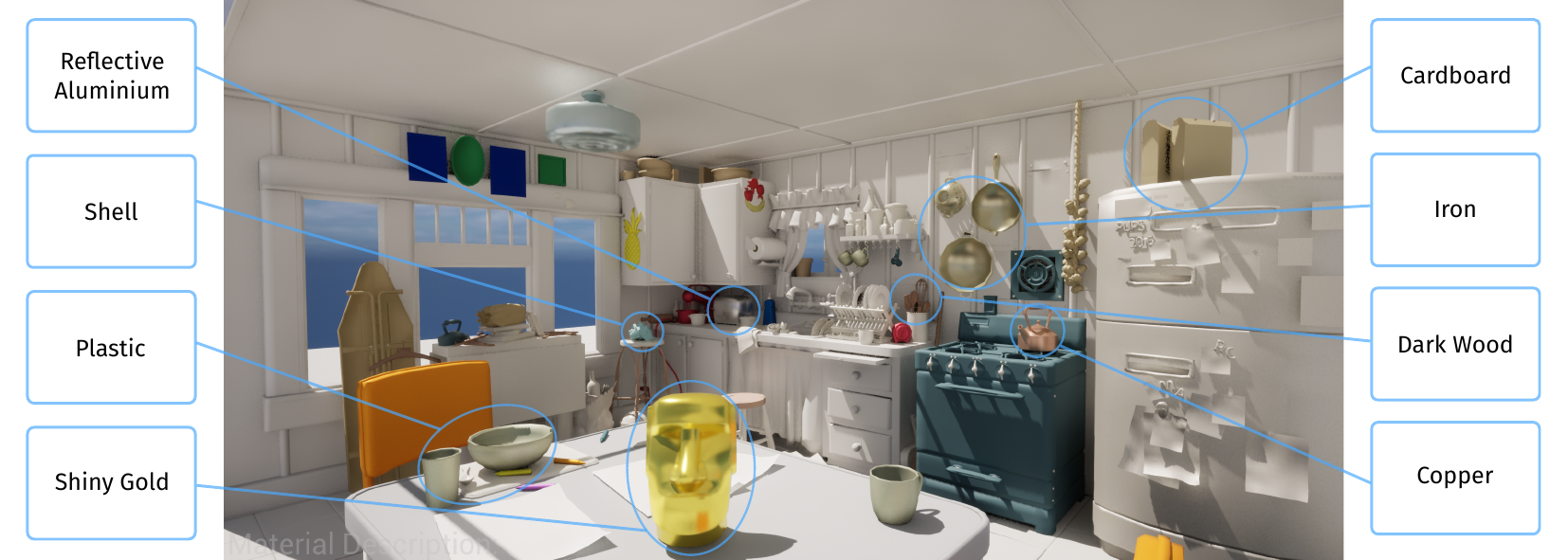}
\caption{\label{fig:SceneImage} \textbf{BRDF Generation from Material Prompts}: Our model allows users to assign materials to objects in real-time scenes by simply describing their desired appearance. This image shows the Kitchen Set USD scene by Pixar, running in Unreal Engine, with materials assigned by typing a description into the text box in the lower left of the screen.}
}
\end{figure*}

\section{Introduction}
Crafting realistic materials for use in physically based rendering (PBR) is a difficult creative process. A highly specialised material designer is often needed for workflows that require accurate rendering of real world objects, e.g. VFX, video games, and visual learning simulations. However, there is a rising trend of user-led design through natural language, i.e. prompting. A user without technical design expertise may wish to simply describe their envisioned material to obtain a BRDF model that matches their description. Although tremendous progress has been made in implicit neural representation of scenes, explicit representation of materials by modeling BRDFs allows tuning and artistic control. In addition, the synthesized material can then be applied to general settings across a variety of geometries, viewing conditions and lighting distributions. In this paper, we address the problem of generation of traditional BRDF material shaders, specifically NVIDIA's Omniverse.

There are two typical representations for PBR materials. Highly accurate measured data can be found from real world objects, tabulating the proportion of light reflected along a set of reflected directions for each discretized incident direction. This format can be highly accurate but has many drawbacks, such as a large memory cost and being difficult to design and interpret. For usage in real-time applications, the BRDF can be estimated using parametric functions known as shaders. In NVIDIA's Omniverse, a recent 3D software platform, the Material Definition Language (MDL) format is used to write shader code specific to representing materials in real-time rendering. Within Omniverse there are multiple preset material definitions that are parametric functions defined by a series of values representing different physical properties. 

Often, for a 3D scene to be created by a designer, a prototype scene is made with fewer time and resources required. Databases of pre-made materials are utilised to speed up development. Our goal is to build a tool to aid in this task by allowing designers, or non-technical users, to describe a scene in text and receive suitable materials. We do this by utilising CLIP \cite{clip}, a multi-modal image and text embedding model. CLIP was trained to encourage embeddings of texts and images to lie close to one another in latent space. In doing so, CLIP provides a correlation rating between text and image pairs, the cosine similarity between the two embeddings, allowing it to be used as a semantic loss during training. With this, we train an autoencoder model to learn a latent space that matches BRDF parameters, conditioned on text embeddings. This architecture can be trained to predict parameters for any target BRDF function in common renderers. Although we demonstrate this with NVIDIA's MDL format, and within a game engine (Unreal Engine), it can equally be applied within the context of other material formats and PBR renderers.

%-------------------------------------------------------------------------
\section{Related Work}
\subsection{BRDF Representation}
As outlined in the survey paper "BRDF Representation and Acquisition" \cite{brdfsurvey}, there are many ways to represent the BRDF. This is due to the many use cases of the function, and the varying complexity of the representations. Fundamentally, a BRDF is a function that takes as input an incoming light direction $\mathbf{v_i}$ and an outgoing direction $\mathbf{v_r}$, and produces the ratio of incoming irradiance to outgoing radiance. The equation is, 

\begin{equation}
    f_r(\mathbf{v_i}, \mathbf{v_r}) = \frac{dL_r(\mathbf{v_r})}{dE_i(\mathbf{v_i})} 
    = \frac{dL_r(\mathbf{v_r})}{L_i(\mathbf{v_i})\cos\mathbf{\theta_i}d\mathbf{v_i}}
\end{equation}

where $E_i$ is the incident irradiance, $L_i$ is incident radiance and $L_r$ is the reflected radiance. For an incident ray colliding with a surface with normal $n$, $\theta_i$ is the measure of the angle between $v_i$ and $n$. In practice however, much of the complexity of the function can be removed as light is simulated simply by measured RGB values, as done in the MERL dataset \cite{Matusik:2003}. MERL stores BRDF values in a tabulated grid of ratios of incoming irradiance to outgoing radiance, used as a multiplier to incoming RGB values. In real-time applications, a more memory-efficient representation is needed, so the BRDF is approximated analytically by shader functions. This is code written in a shader language, such as GLSL \cite{glsl}, to be run in parallel on the GPU. In NVIDIA Omniverse, there are many shader functions that can be used to simulate different material types, all represented in the Material Definition Language (MDL). OmniPBR is the default material, it can describe dielectric and non-dielectric materials, but lacks features such as sub-surface scattering. OmniSurface is a more intricate BRDF implementation that can represent a wider range of materials.

Neural representations of measured BRDFs have become more popular in recent years for many of the same reasons analytical BRDFs are popular \cite{sztrajman_neural_2021}\cite{rainer_neural_2019}. Dimensionality reduction of measured BRDFs \cite{tongbuasirilai_sparse_2022}\cite{chen_invertible_nodate}\cite{zheng_compact_2022} is particularly useful as it brings many desirable attributes such as lower memory usage, faster parsing, and more human readable values. Similarly, learning a latent space of dimensionality reduced representations is useful for design and material interpolation \cite{benamira_interpretable_2022}\cite{hu_deepbrdf_2020}. 

Another longstanding challenge in graphics programming has been BRDF estimation from other modalities. The most common form of this is to estimate BRDF values from images of real world materials. These images often consists of single photographs, possibly captured with flash. To achieve a successful estimation, it is necessary to render a visually similar image by generating normal, roughness, and metallic maps. Methods face numerous obstacles, including complications arising from camera flash and the accurate representation of fine details \cite{rhee_estimating_2022}\cite{boss_single_2019}\cite{otani_brdf_2021}.\cite{zhou2023photomat} introduces a material generator learned from single images of surfaces.

\subsection{Generation from Text}
With the coming of more effective language modelling methods, there has been an explosion of work on generating 3D content from text. Many of this has been directly inspired from other modalities, such as diffusion methods as used in 2D image generation, applied to 3D models \cite{huang_joint_2023}. Other work has focused on general scene generation, either unconditioned \cite{renderdiffusion} or conditioned on user data, such as object positioning for example \cite{po_compositional_2023}. The majority of these methods focus on implicit object generation, that is predicting objects whose shape and appearance are contained in the neural model that generates them. This is best shown by the popularity of Neural Radiance Fields \cite{mildenhall}\cite{lin_componerf_2023}. NeRFs are neural representations of scenes that are entirely contained in the network, with no explicit graphics resources required. There has been a small amount of work done on explicit generation from text, where the generated content is separate from the model that created it \cite{khalid_clip-mesh_2022}\cite{chen_fantasia3d_2023}. In particular there is Point-E \cite{nichol2022pointe}, a 3D point cloud generation technique that is conditioned on text prompts. It utilises text to 2D image generation and image to point cloud diffusion, while being guided by text prompts. There is a focus on efficiency over accuracy in this work, showing promise for an eventual real-time application. There is also Fantasia3D \cite{chen2023fantasia3d}, a recent work that learns to predict geometry and appearance from text prompts, that can generate multiple texture maps of a 
small set of microfacet BRDF parameters. The results are a highly detailed Spatially-Varying BRDF (SVBRDF) that is trained for each material prompt.

Recent years have seen a rise in work relating text and visuals, mirroring the rise in popularity of generative text models. \cite{wu2020describing} present an early work in generating descriptions and gives a thorough study of the available methods. There has also been multiple concurrent works that generate materials conditioned on text prompts. \cite{Hu_2023} utilise an extensive material dataset to generate high quality material node graphs using an autoregressive model conditioned on a CLIP text or image embedding. \cite{deschaintre23_visual_fabric} use a large dataset of fabric materials and descriptions to fine-tune CLIP for material retrieval. Both works make use of CLIP, showing a growing trend within data-driven material generation.

\subsection{Stylising from Text}
A popular use of natural language conditioning on 3D content is stylisation. This is the editing of 3D content to match a given style, as described in a text. A popular recent method for this is the use of CLIP embeddings as a semantic loss during training. \cite{michel_text2mesh_2022} propose Text2Mesh, a model for estimating vertex colour and displacement values in a given mesh, so as to fit a style described in text. While this is effective at the chosen task, the edited meshes still lack many modern graphics components. \cite{chen_tango_2022} improve on the visualisation of Text2Mesh by estimating a wider range of material properties, to achieve a more realistic rendering of the given mesh. Both of these works however represent their graphics content implicitly within their models, making their methods difficult to integrate into traditional workflows. TANGO proposes disentangling their neural representation and exporting it as explicit graphics resources, which is effective but not efficient enough to be feasible for large amounts of data.

%-------------------------------------------------------------------------
\section{Method}
To develop a real-time material generation system, several components are required. First is an efficient physically based renderer that can provide accurate images of generated materials, this is NVIDIA Omniverse. Second is a method of evaluating the generated images against the input text, this is the semantic loss from CLIP. Third is a source of training data, in this case text and material pairs. Lastly is a machine learning model that can make material parameter predictions from the CLIP text embedding inputs. In our
system, this is a simple fully connected autoencoder model.

\subsection{Data}
\label{Data}
Training the model requires many samples of text that materials can be generated from. There were two sources of this data, the first is from several popular websites of collections of materials \cite{matweb1}\cite{matweb2}. We simply found the 100 most common nouns and 25 most common adjectives from the material names across each website, with table \ref{NounsAdjs} displaying the 10 most common of each. We combined these these to create a dataset of material prompts with either 0 or 1 adjectives followed by a noun. 

The second source of material prompts is from several publicly available novels, all nouns and associated adjectives are parsed using part of speech tagging with the NLTK library \cite{NLTK}. The parsed descriptions are then screened using concreteness ratings available online \cite{concreteness}, where descriptions are only saved if their noun is found to have a particularly high concreteness rating (at least 4.5 from a maximum of 5). This method can provide realistic material prompts with multiple adjectives, rather than using random combinations of adjectives from source one, which may contradict each other e.g. "wet, dry rock".

\begin{center}
\begin{tabular}{ |c|c| } 
\hline
Nouns & Adjectives \\
\hline
wood & metallic \\
metal & shiny \\
brick & smooth \\
concrete & polished \\
stone & rough \\
glass & matte \\
marble & reflective \\
gold & glossy \\
plastic & dull \\
leather & dark \\
\hline
\end{tabular}
\captionof{table}{10 most common material nouns and adjectives.}
\label{NounsAdjs}
\end{center}

\subsection{Training}
\begin{figure*}[h]
\centering 
{
\includegraphics[width=1.05\linewidth]{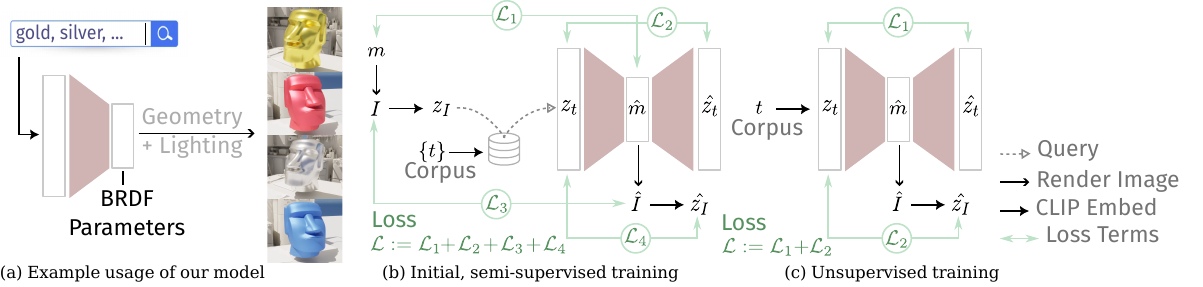}
\caption{\label{fig:ModelTraining} Overview of our training process, both semi-supervised and unsupervised learning.}
}
\end{figure*}
There are two methods for training the autoencoder model, based on the two sources of material prompts, semi-supervised and unsupervised. 
The semi-supervised training procedure requires pairs of material and material descriptions. To get these, we utilise CLIP once more to annotate rendered images of randomly generated materials. To do this we generate a random vector of material parameters $m$, and render an image $I$ of a sphere with this material applied to it. A CLIP image embedding $z_I$ is found from the material render, and a vector database of CLIP text embeddings is queried. This vector database is made from the material prompts created from common nouns and adjectives, described in \ref{Data}. With this method we can find the material prompt embedding $z_T$ that best fits the material render. These values make up a single semi-supervised input for the model. 

The material prompt embedding $z_T$ is given as input to the autoencoder model. A vector of material parameters $\hat{m}$ is found as the model's latent representation of the input embedding. The output of the model is the reconstructed text embedding $\hat{z}_T$. During training, the predicted vector of material parameters $\hat{m}$ is rendered in the same environment as the target material to get the reconstructed material image $\hat{I}$, which is then given to CLIP to calculate the final image embedding $\hat{z}_I$. Finally, all values have been calculated for the semi-supervised loss function:
\begin{equation}
    L = || \hat{m}, m||_1 + || \hat{z}_T, z_T ||_1 + || \hat{z}_I, z_I ||_1 + CLIP_{sim}(z_T, \hat{z}_I, z_I)
\end{equation}
Where the $CLIP_{sim}$ function is one minus the difference between the cosine similarities of the prediction and target embeddings:
\begin{equation}
    CLIP_{sim}(z_T, \hat{z}_I, z_I) = 1 - (sim(z_T, \hat{z}_I) - sim(z_T, z_I))
\end{equation}

The model is then further trained in an unsupervised manner. Using the text from novels and concreteness screening, a material prompt is chosen randomly. This prompt is embedded by CLIP, as $z_T$, and given to the model as input. Identical to the semi-supervised training, a vector of material parameters $\hat{m}$ and a reconstructed text embedding $\hat{z}_T$ are outputted by the model. The CLIP image embedding of the rendered material, $\hat{z}_I$, is also generated as in the semi-supervised training. These are the values used to calculate the unsupervised loss:
\begin{equation}
    L = || \hat{z}_T, z_T ||_1 + sim(z_T, \hat{z}_I) 
\end{equation}

\subsection{Material Post-Processing} \label{postprocessing}
Along with the material parameters found within the latent representation, there are also others which were chosen to be predicted in another way, due to nature of the visual learning task. For example, though it may seem natural to predict the diffuse colour of the material along with the other material parameters, we found that this caused the model to ignore other visual features and only attempt to predict colour. This is likely due to the larger influence colour has than other material parameters on the final rendered image. As a simple solution to this, we predict colour, along with transparency and refractive index, by querying a vector database of text and value pairs, to find the most similar text to the material prompt and assign the paired value. A vector database is a collection of text embeddings, each with saved values, that allows querying based on a nearest neighbour search i.e. the nearest embeddings to a search embedding are found and their saved values retrieved. We use the open source library Chroma \cite{chroma} for this project. 

To collect the text-value pairs, we used multiple text-colour sources \cite{wikicolour}\cite{xkcdcolours}, refractive index tables \cite{refradctiveindex}, and transparency values from material examples \cite{omniversematerials}. Importantly, only a relatively small amount of data is needed for this task as a weighted sum of values can be calculated using the text embedding distances found during querying. This allows diversity within the queries without needing extensive databases of values. 
% Figure \ref{VDB} is a visualisation of each of the vector databases, for colour, refractive index, and transparency.

\subsection{Implementation}
In our implementation, the autoencoder model has $5$ encoder layers and $5$ decoder layers. The first layer has a size of $512$, matching the dimension of the CLIP text embedding. Each subsequent layer is half the size of the previous, until the fifth layer with a size of $32$. The decoder layers mirror those in the encoder. The latent space is the size of the chosen BRDF parameters, in this case $8$ for the parameters of the OmniSurface function in NVIDIA's MDL format. Every layer is fed into a ReLU connection and the final decoder layer feeds into a sigmoid function. The model was trained with the Adam optimizer and a learning rate of 1e-4, for several thousand training steps. Our implementation was trained using an RTX 4090 with 24GBs of VRAM, enabling the training of the model and the rendering of images to be done simultaneously. We rendered images using NVIDIA Omniverse's Python API which was called within a Pytorch training loop. Renders used in training were of a simple scene of a sphere with the current material applied and a bright environment map. The images were cropped to remove all visuals but the sphere, to allow CLIP to focus on the material.

%-------------------------------------------------------------------------
\section{Evaluation}
\subsection{Survey}
We evaluated our model via a survey of 125 users on Amazon Mechanical Turk who indicate pairwise preference for a given prompt. As a comparison to images rendered with predictions made by our model, we used the Base Materials pack available on NVIDIA Omniverse. This is a database of common materials such as "Beige Carpet", "Concrete Rough", "Brick", etc. They are created using the "OmniPBR" MDL function to estimate the BRDF, this is a default function of Omniverse and capable of realistically representing diverse materials. Our survey was presented as a series of pairs of images with a single text input which we call the material prompt. We constructed a fixed set of 250 possible prompts by combining the 25 most common nouns and 10 most common adjectives in our text corpus. The target material from the Base Materials pack was chosen as the material whose name is found to be closest to the prompt, such as "Concrete Rough" for the prompt "Concrete". This is in the spirit of what a designer might choose given  a material description. The second image in the pair is a render of an identical scene, but with a material generated from our model with the material prompt as input. The survey-taker either chooses their preferred image for each prompt or may click "Unsure" if they are unable to decide. Pairs were shown in random order and sometimes repeated, after flipping, within a survey.

\begin{figure*}[h]
\begin{tabular}{@{}c@{}c@{}c@{}}
\includegraphics[width=.358\linewidth]{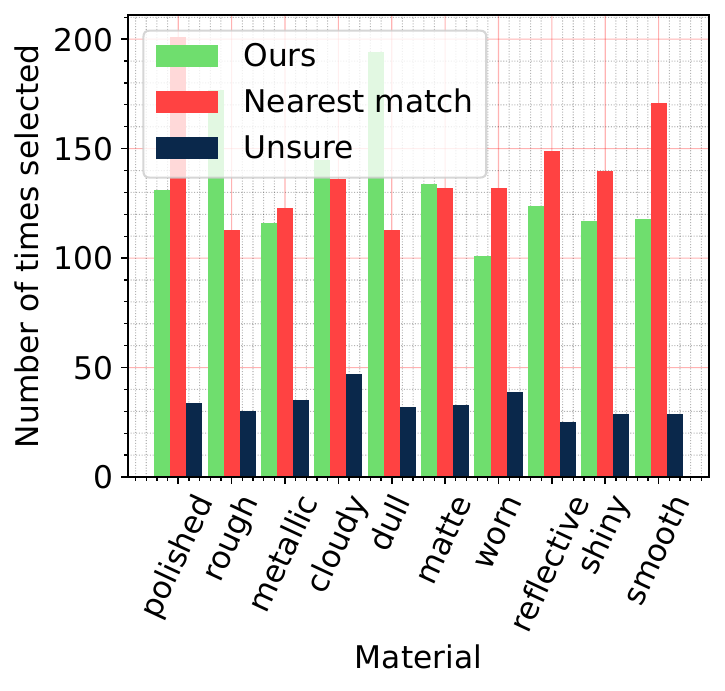}&
\includegraphics[width=.61\linewidth]{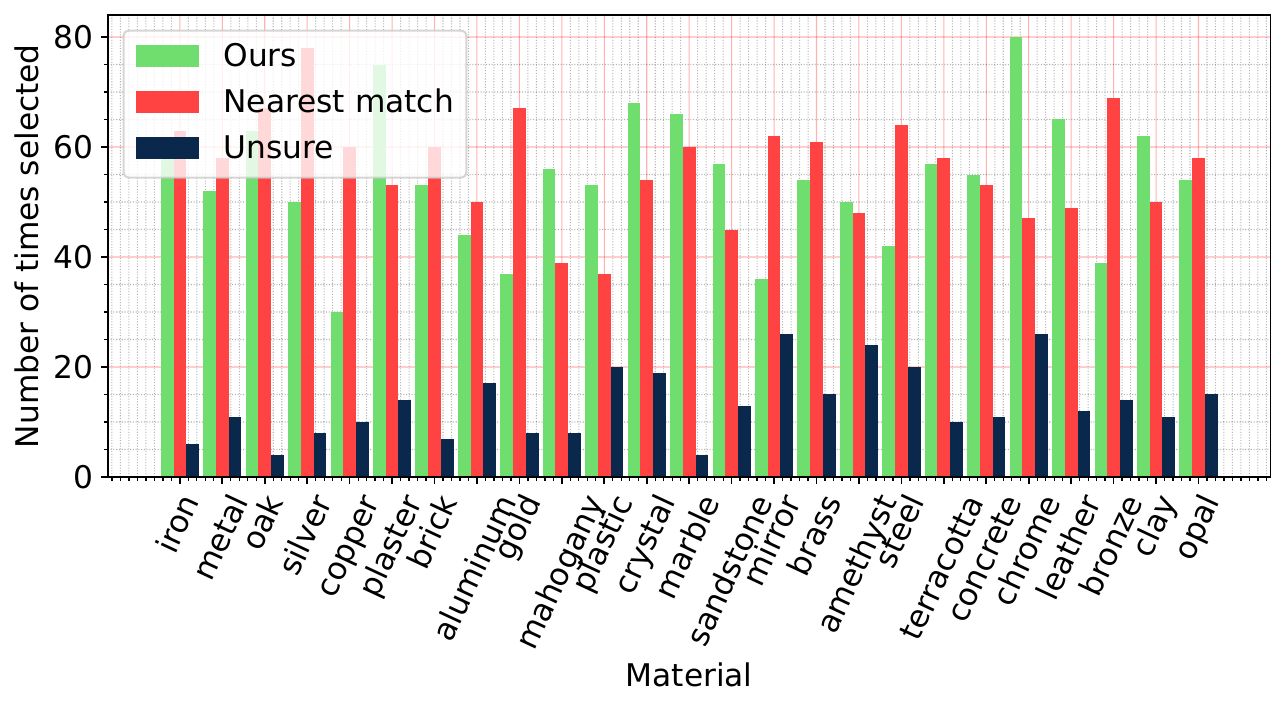}
\end{tabular}
\caption{Results from our survey. The plots show the number of times users chose an image rendered with our (green) BRDF against an image rendered using a BRDF from NVIDIA's Omniverse database, for a given text prompt. The plots show users' preferences across adjectives (a) and nouns (b).}
\label{survey}
\end{figure*}

\begin{figure}[h]
\begin{center}
\includegraphics[width=.8\linewidth]{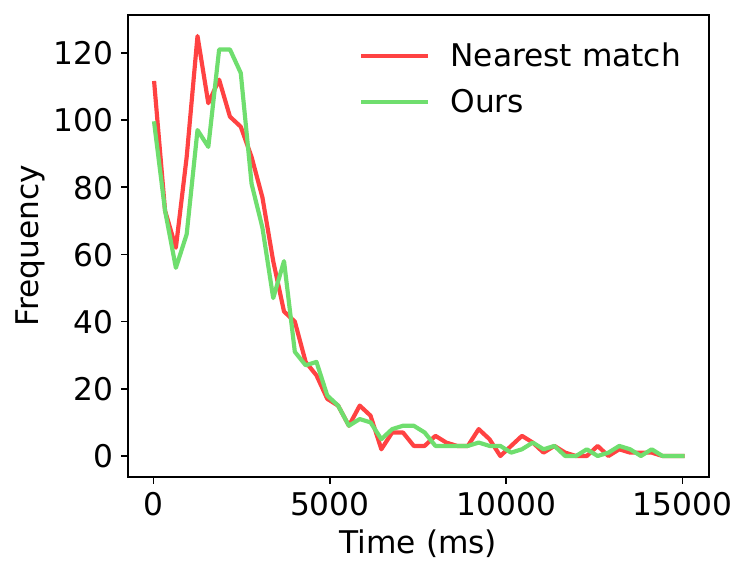}
\caption{\label{fig:surveytime} Histograms of user response times in the survey when selecting our BRDF (green) vs a material from the database (red). The histograms are virtually indistinguishable. The slightly larger mode for our histogram suggests a subtle trend showing that users who took longer could have had a  preference for our BRDFs. }
\end{center}
\end{figure}

Figure \ref{survey} summarises the results of our survey. It shows that on average, users were unable to distinguish BRDFs generated by our model from BRDFs in the database. However, the model seemed to perform better for some materials (such as chrome) than others (such as gold). Figure~\ref{fig:surveytime} shows the histogram of response times by users when they chose images rendered with our materials (green) vs those rendered with images from the Base material database. The summary of these results is that materials automatically generated by our model were indistinguishable from those from the manually-crafted database associated with a specific prompt.

\subsection{Example Applications}
\begin{figure*}[!htb]
    \centering
    \begin{subfigure}[b]{0.33\textwidth}
        \includegraphics[width=\textwidth]{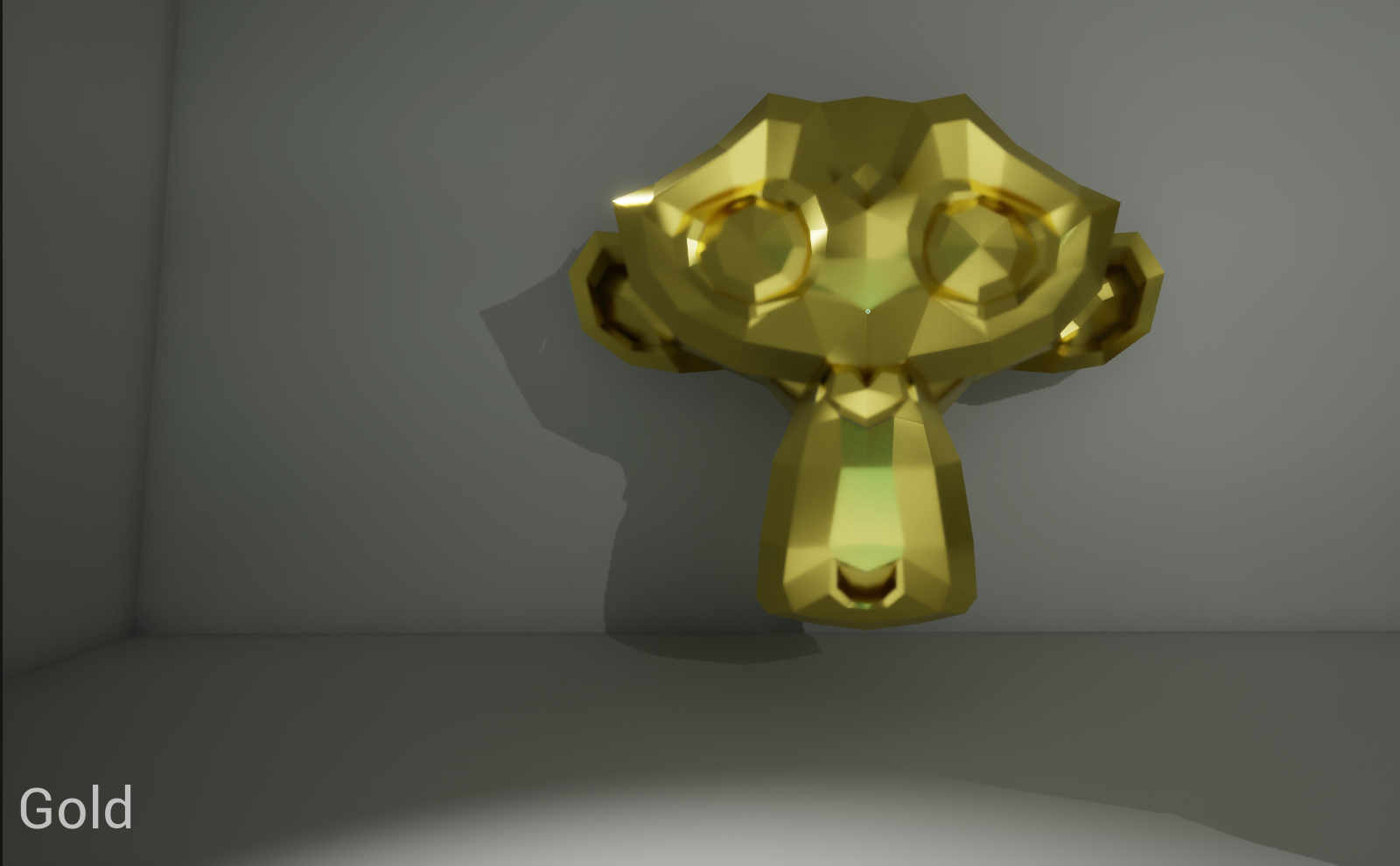}
        \caption{Material Prompt: \textbf{Gold}}
        \label{fig:image1}
    \end{subfigure}
    \hfill
    \begin{subfigure}[b]{0.33\textwidth}
        \includegraphics[width=\textwidth]{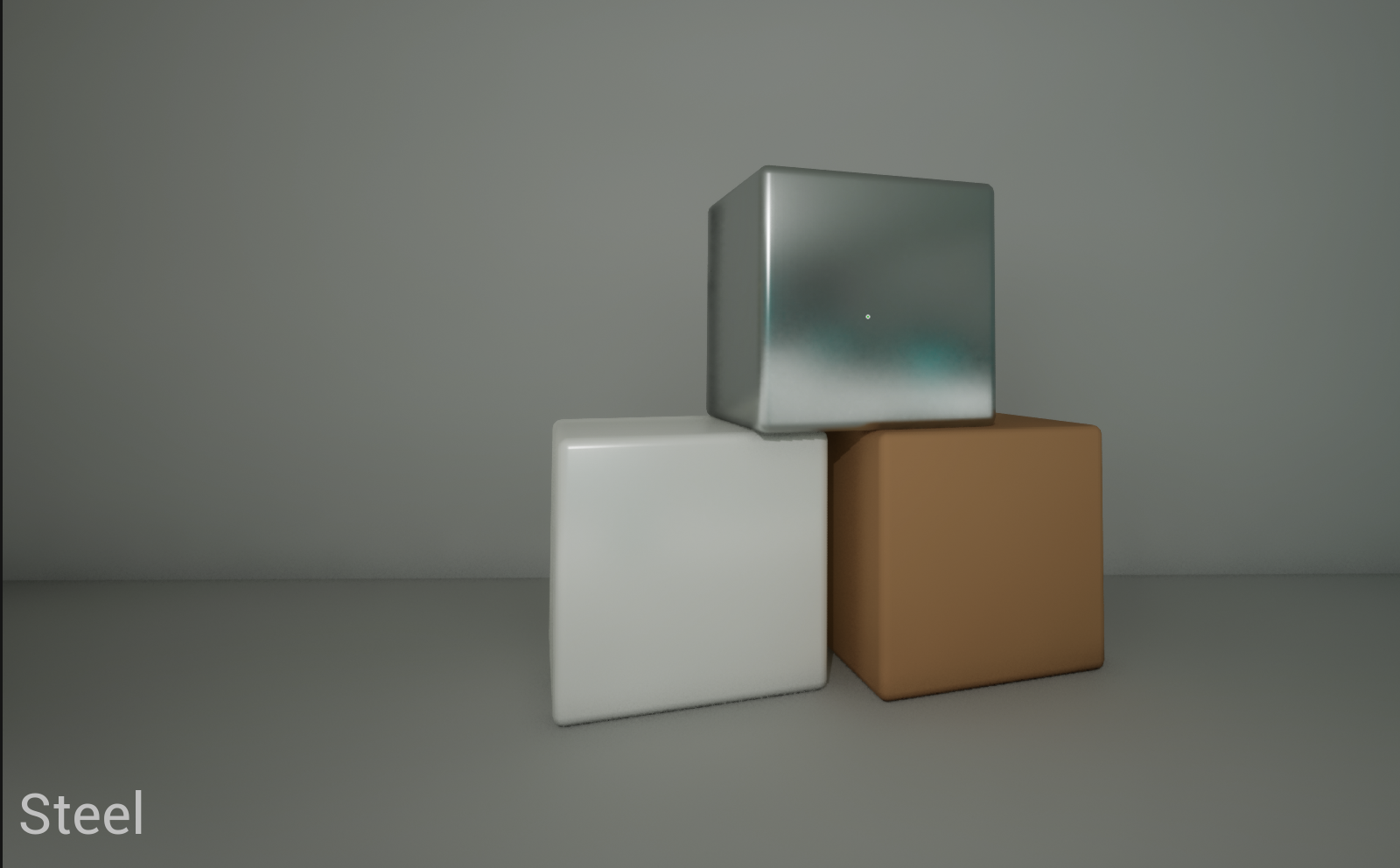}
        \caption{Material Prompt: \textbf{Steel, Plastic, Brick}}
        \label{fig:image2}
    \end{subfigure}
    \hfill
    \begin{subfigure}[b]{0.33\textwidth}
        \includegraphics[width=\textwidth]{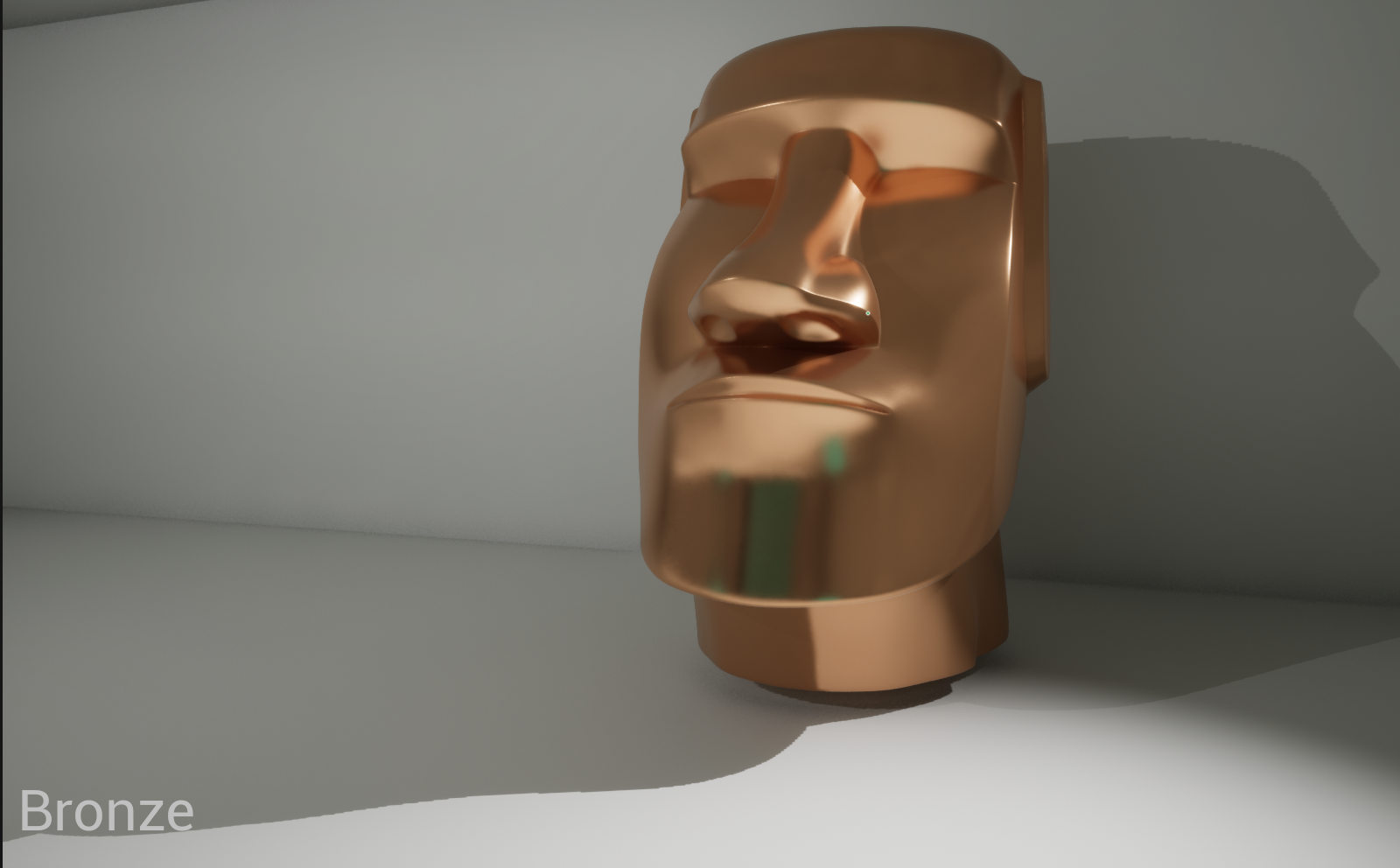}
        \caption{Material Prompt: \textbf{Bronze}}
        \label{fig:image3}
    \end{subfigure}
    \caption{Examples from the Unreal Engine demo of our model.}
    \label{Unreal}
\end{figure*}

To demonstrate practical use of the model, we developed applications for two possible use cases. The first is an Unreal Engine integration allowing a player to type a material prompt and simply click to select an object in the world to apply a predicted material onto. Screenshots of this are shown in figure \ref{Unreal}, with multiple example prompts. It's important to note that due to the size of the model, inference can be extremely fast. Encoding a text embedding to a material vector takes just 5ms in our tests, allowing our Unreal Engine demo to set materials in real-time.

\begin{figure*}[!htb]
    \centering
    \begin{subfigure}[b]{0.3\textwidth}
        \includegraphics[width=\textwidth]{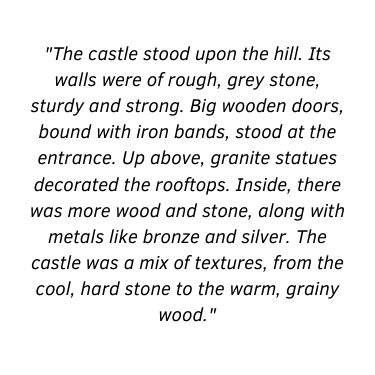}
        \caption{Scene description.}
        \label{fig:image1}
    \end{subfigure}
    \hfill
    \begin{subfigure}[b]{0.3\textwidth}
        \includegraphics[width=\textwidth]{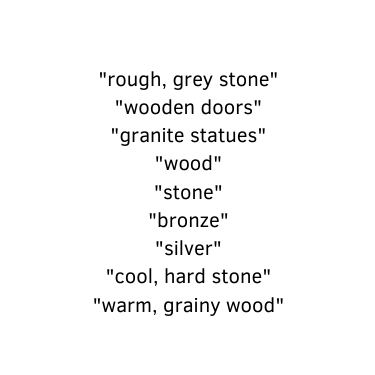}
        \caption{Parsed material prompts.}
        \label{fig:image2}
    \end{subfigure}
    \hfill
    \begin{subfigure}[b]{0.3\textwidth}
        \includegraphics[width=\textwidth]{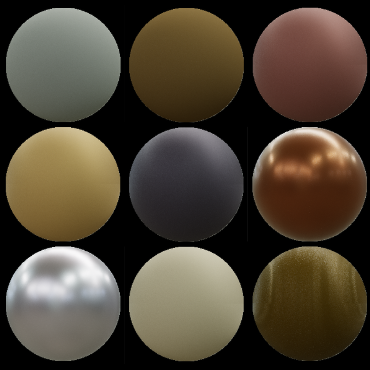}
        \caption{Collection of generated materials.}
        \label{fig:image3}
    \end{subfigure}
    \caption{An example scene initialisation utilising our model.}
    \label{SceneInit}
\end{figure*}

Our second demo is a system for initialising a scene conditioned on a paragraph description. Our application parses the paragraph of text and extracts the nouns and their associated adjectives. These are then screened for concreteness using publicly available concreteness ratings of words \cite{concreteness}. The final material prompts are then inputted to the model and a collection of material vectors are generated. These vectors of parameters are then used to edit a default MDL file, with the target BRDF function, to export the materials as files. These material files can then be used within software such as NVIDIA Omniverse to initialise a scene. Figure \ref{SceneInit} shows a visualisation of the process. 

\subsection{Word2Vec Comparison}
\begin{figure}[!htb]
\includegraphics[width=8cm]{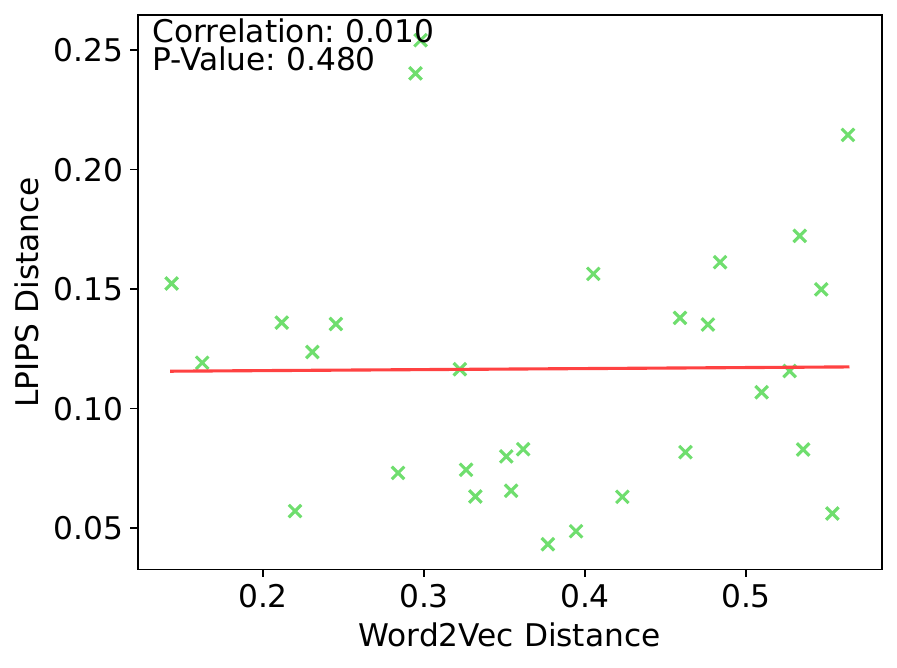}
\caption{Correlation of LPIPS distance and Word2Vec distance, for full colour images.}
\label{LPIPS}
\includegraphics[width=8cm]{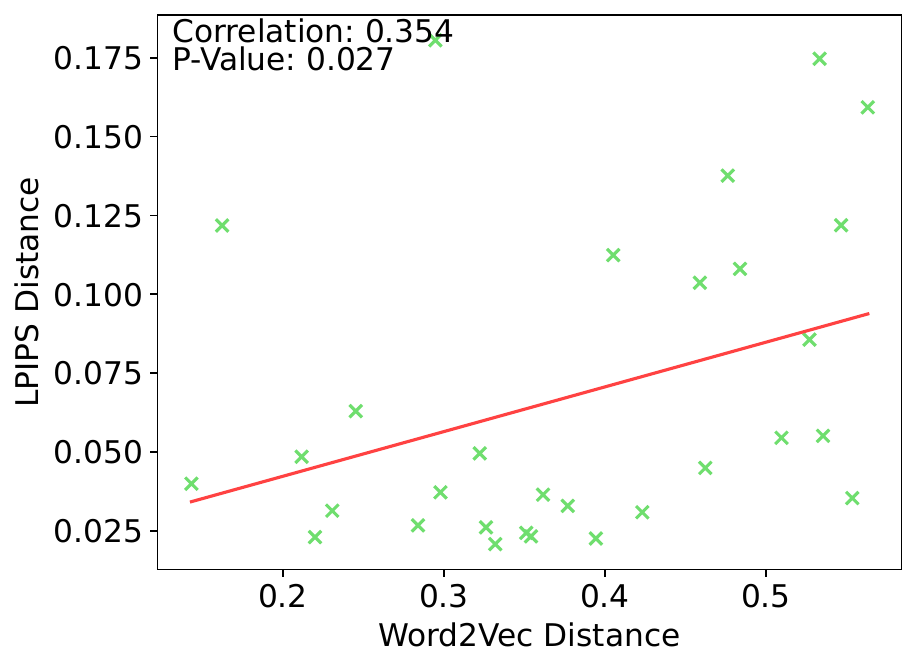}
\caption{Correlation of LPIPS distance and Word2Vec distance, for grey images.}
\label{LPIPS_grey}
\end{figure}
To evaluate the material latent space learned by our model, we compare the distribution of samples with two metrics: LPIPS \cite{LPIPS} distance metric, and Word2Vec \cite{w2v} distance. LPIPS is a similarity metric between images that utilises the final layers of deep neural networks, we use it to find the similarity between predicted material samples and the Word2Vec neighbours of their material prompts. To study these distributions, we get the 10 most common adjectives from our semi-supervised dataset and find the 10 nearest neighbours of their Word2Vec representations. Note that we remove the antonyms from the set of neighbours as Word2Vec is known to group words with inverse meanings. We use these neighbours, along with the 25 most common nouns from our dataset, to create material prompts for our model and get the LPIPS distance between rendered images of the neighbour prompt materials and starting prompts. 

In figure \ref{LPIPS}, we plot the Word2Vec distance of each material prompt to its starting prompt compared to its LPIPS distance to the starting prompt. This plot compares distances of fully coloured materials, as described in section \ref{postprocessing}, which is likely over represented in the LPIPS similarity metric. To study the effects of only the material parameters on the correlation of distances, figure \ref{LPIPS_grey} shows the same experiment but all predicted materials use the same grey base colour. This forces LPIPS to only vary based on the material parameters and not colour.  

\subsection{Interpolation}
\begin{figure}[!htb]
\includegraphics[width=8cm]{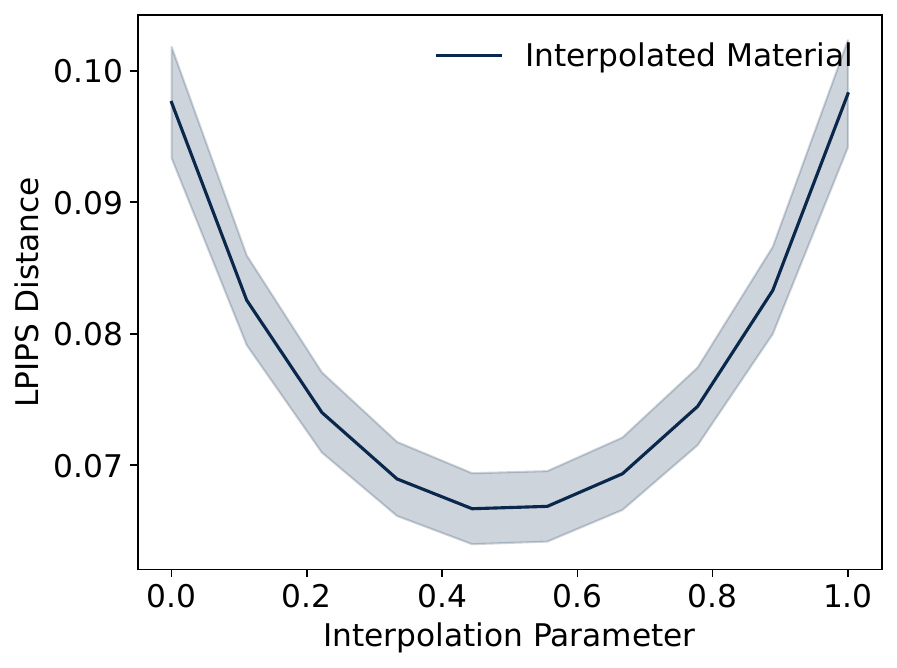}
\caption{LPIPS distance between the interpolations of predicted and ground truth materials.}
\label{interpolation1}
\end{figure}
\begin{figure}[!htb]
\includegraphics[width=8cm]{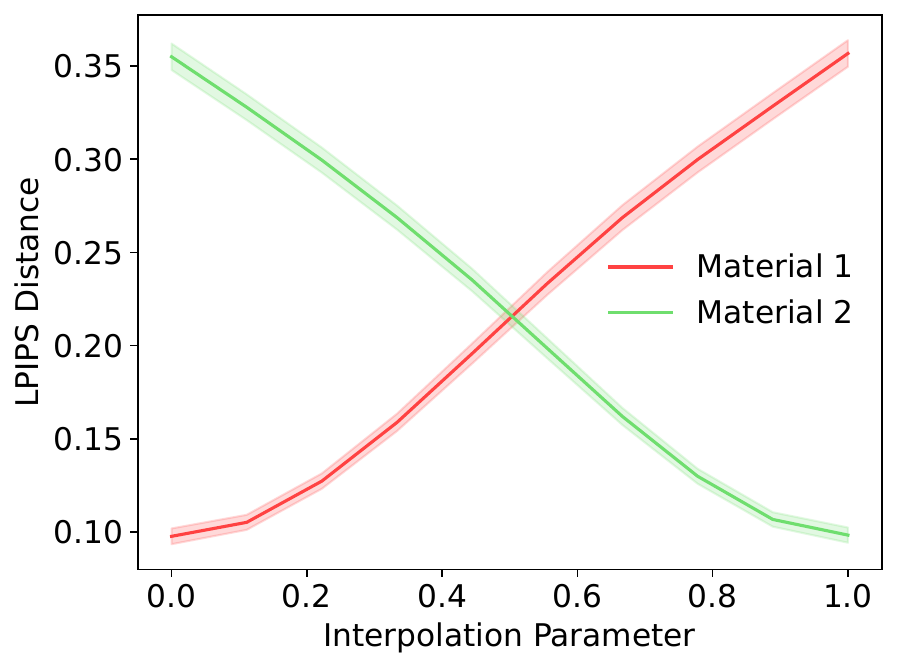}
\caption{Interpolation of predicted materials and LPIPS distance to fixed ground truths.}
\label{interpolation2}
\end{figure}
Figures \ref{interpolation1} and \ref{interpolation2} show our study of the interpolation of our model's BRDF latent space. Both plots are generated from 1000 samples. 

In figure \ref{interpolation1}, two pairs of ground truth material and text are selected, then two predicted materials are generated using the ground truth texts. The vectors of material parameters are then interpolated, between the first ground truth material and the second, and between the first predicted material and the second. Finally, the LPIPS distance between rendered images of the two interpolated materials are plotted. 

In figure \ref{interpolation2}, two ground truth material and text pairs are also selected, with predicted materials generated similarly. In this case, only the predicted materials are interpolated, between the first prediction and the second. The LPIPS distance between the interpolated material and the first ground truth material is plotted in blue, and the distance to the second ground truth material is plotted in orange. 

%------------------------------------------------------------------------

\section{Discussion}

\subsection{Word2Vec Comparison}
Figures \ref{LPIPS} and \ref{LPIPS_grey} show the results of our study of the latent space of our model. The correlation in the first figure is near zero, with a high p-value, implying there is no relation between Word2Vec distance and LPIPS distance, however this is only for the full colour renders of materials. Figure \ref{LPIPS_grey} compares the two metrics on renders of materials with a grey base colour, and shows a positive correlation coefficient of $0.354$ with a low p-value of $0.027$. This suggests that an increase in the Word2Vec distance between two prompts relates to an increase in the LPIPS distance between the rendered images of the prompts predicted materials. As we remove antonyms from the neighbour set, this result implies that the predictions made by our model become visually dissimilar as the material prompts' meanings become more dissimilar. As this correlation is present only when materials lack post-processing (i.e. base colour), it suggests that our model follows a similar distribution to Word2Vec, but that LPIPS is highly sensitive to material colour.

\subsection{Interpolation}
Our two interpolation results have two different conclusions. Figure \ref{interpolation1} shows the interpolation of predicted and ground truth materials. We see a significant decrease in the distance between the two as interpolation occurs, before returning to the initial distance. This is to be expected within a smooth space as for both interpolations the beginning and end states are "near" each other, as our model is trained to make nearby predictions. Therefore, on average, the interpolated materials will approach each other as interpolation increases, before diverging again as interpolation finishes. 

The second plot, figure \ref{interpolation2}, has a very clear interpretation. Predictions begin near their ground truth targets (the blue line begins low) before finishing far from their targets after being interpolated to another material (the blue line ends high). The inverse of this is shown in the orange line which lowers as the interpolation approaches the orange line's ground truth material.

Both of these results imply a smooth material representation that acts predictably under interpolation.

%------------------------------------------------------------------------

\section{Limitations}
\subsection{Textual Prompt Space}
Although the text prompts (adjective(s) + noun) used to generate materials are relatively simple, they are actually not trivial to work with. In fact, having multiple adjectives within the same text prompt can lead to incongruent results due to contradictory or even vague adjectives. Furthermore, CLIP model struggles to annotate randomly generated materials with text prompts possessing multiple adjectives. To overcome these problems, a two-phase approach was implemented, starting with a semi-supervised training with single adjectives followed by an unsupervised training on text prompts with multiple adjectives. Richer textual prompts were out of the scope of this paper due to the small amount of labelled data (material/prompt) available and the difficulty to obtain high quality materials.

\subsection{Learning BRDF via Abstract Material Descriptions}
Since the focus of the encoder is not to learn a neural representation of a BRDF but to learn the parameters of various materials instead, using an abstract material description such as MDL for describing properties (surface, physically based, glass-with-volume, etc.) was a straightforward choice. Moreover, MDL is a shading language that does not produce programs for a particular renderer but rather it defines the behaviour of light at a higher level, which makes it a suitable material description for working with different renderers and under different settings (end-user applications or within more complex workflows). However, it is important to take into account the practical implications of the chosen material description, e.g. rendering time, complexity and portability, as they can easily hinder training and applications of the generated materials. As an anecdote, we tried to use our method with PBRT material descriptions, but the high rendering time became a bottleneck while training and ultimately had to be abandoned.

\subsection{Material Post-Processing}
Material generation involves learning complex, orthogonal concepts/spaces such as material properties, colour, texture map, normal map, displacement maps, etc. Whether we should frame this problem as a group of independent learning tasks or rather as a combined single task is unclear, the literature has adopted different approaches based on the scope, resources and limitations of the project. Besides, since the project is aimed to end-users, it is important to take into account how many different interpretations of a prompt are possible and therefore, it is quite likely that properties such as colour, transparency and refractive index of the generated materials will be further tuned to fit user needs.
%------------------------------------------------------------------------
\section{Conclusion}
In this paper, we presented a method for training an autoencoder model to learn a latent space of BRDF parameters. The model can be implemented to predict parameters for any parametric BRDF implementation, with our work utilising the MDL functions available in NVIDIA Omniverse for training data and evaluation. Our demos show the practical use of such a model, and the capability of real-time generation in Unreal Engine. We studied how the latent space of our model represents BRDF parameters through interpolation experiments and comparisons with the word distribution in Word2Vec. We believe the results of our evaluations show an effectiveness in our training method, that a semantic loss for text and rendered images is capable of learning subtle parameters such as material values. However, our survey shows the extent of this method, as it was found to be indistinguishable but not surpassing database retrieval.

%-------------------------------------------------------------------------

% Bibliography
\newpage
\bibliographystyle{ACM-Reference-Format}
\bibliography{egbibsample}

%-------------------------------------------------------------------------

\end{document}